\documentclass[12]{article}
\usepackage{fleqn}
\usepackage{graphicx}
\begin{document}
\Large
\begin{center}
{\bf  Hjelmslev Geometry of Mutually Unbiased Bases}
\end{center}
\vspace*{-.1cm}
\begin{center}
Metod Saniga$^{\dag}$ and Michel Planat$^{\ddag}$
\end{center}
\vspace*{-.1cm} \normalsize
\begin{center}
$^{\dag}$Astronomical Institute, Slovak Academy of Sciences,\\
05960 Tatransk\' a Lomnica, Slovak Republic\\
(msaniga@astro.sk)

\vspace*{.1cm}
 and

\vspace*{.1cm} $^{\ddag}$Institut FEMTO-ST, CNRS, Laboratoire
de Physique et M\'etrologie des Oscillateurs,\\ 32 Avenue de
l'Observatoire, F-25044 Besan\c con, France\\
(planat@lpmo.edu)
\end{center}

\vspace*{.0cm} \noindent \hrulefill

\vspace*{.0cm} \noindent {\bf Abstract}

\noindent
The basic combinatorial properties of a complete set of
mutually unbiased bases (MUBs) of a $q$-dimensional Hilbert space
${\cal H}_q$, $q = p^{r}$ with $p$ being a prime and $r$ a
positive integer, are shown to be qualitatively mimicked by the
configuration of points lying on a proper conic in a projective
Hjelmslev plane defined over a Galois ring of characteristic $p^2$
and rank $r$. The $q$ vectors of a basis of ${\cal H}_q$
correspond to the $q$ points of a (so-called) neighbour class and
the $q+1$ MUBs answer to the total number of (pairwise disjoint)
neighbour classes on the conic.

\vspace*{.15cm} \noindent {\bf MSC Codes:} 51C05 -- 81R99 -- 81Q99

\vspace*{.05cm} \noindent {\bf PACS Numbers:} 02.10.Hh -- 02.40.Dr -- 03.65.Ca

\vspace*{.05cm} \noindent {\bf Keywords:} Projective Hjelmslev Planes -- Proper Conics --
Mutually Unbiased Bases

\vspace*{-.1cm} \noindent \hrulefill

\vspace*{.3cm}  \noindent Two distinct orthonormal bases of a
$q$-dimensional Hilbert space, ${\cal H}_{q}$, are said to be
mutually unbiased if all inner products between any element of the
first basis and any element of the second basis are of the same
value $1/\sqrt{q}$. This concept plays a key role in a search for
a rigorous formulation of quantum complementarity and lends itself
to numerous applications in quantum information theory. It is a
well-known fact (see, e.g., \cite{pr05}--\cite{spr} and references
therein) that ${\cal H}_{q}$ supports at most $q+1$ pairwise
mutually unbiased bases (MUBs) and various algebraic geometrical
constructions of such $q+1$, or {\it complete}, sets of MUBs have
been found when $q = p^{r}$, with $p$ being a prime and $r$ a
positive integer. In our recent paper \cite{sp05} we have
demonstrated that the bases of such a set can be viewed as points
of a proper conic (or, more generally, of an oval) in a projective
plane of order $q$. In this short note we extend and qualitatively
finalize this picture by showing that also individual vectors of
every such a basis can be represented by points, although these
points are of a different nature and require a more general
projective setting, that of a projective {\it Hjelmslev} plane
\cite{hj23}--\cite{dem68}.

To this end in view, we shall first introduce the basics of the Galois ring theory (see, e.g., \cite{wan} for the symbols, notation and further
details).
Let, as above, $p$ be a prime number and
$r$ a positive integer, and let $f(x) \in {\cal Z}_{p^2}[x]$ be a monic polynomial of degree $r$ whose image in ${\cal Z}_{p}[x]$ is irreducible. Then
$GR(p^{2},r) \equiv {\cal Z}_{p^2}[x]/(f)$ is a ring, called a {\it Galois ring}, of characteristic $p^2$ and rank $r$, whose maximal ideal is
$p\,GR(p^{2},r)$. In this ring there exists a non-zero element $\zeta$ of order $p^{r}-1$ that is a root of $f(x)$ over ${\cal Z}_{p^2}$, with
$f(x)$ dividing $x^{p^{r}-1}-1$ in ${\cal Z}_{p^2}[x]$. Then any element of $GR(p^{2},r)$ can uniquely be written in the form
\begin{equation}
g = a + p b,
\end{equation}
where both $a$ and $b$ belong to the so-called Teichm\" uller set ${\cal T}_{r}$,
\begin{equation}
{\cal T}_{r} \equiv \left\{0, 1, \zeta, \zeta^{2},\ldots , \zeta^{p^{r}-2} \right\},
\end{equation}
having
\begin{equation}
q = p^{r}
\end{equation}
elements. From Eq.\,(1) it is obvious that $g$ is a unit (i.e., an invertible element) of $GR(p^{2},r)$ iff $a \neq 0$ and a zero-divisor
iff $a = 0$. It then follows that $GR(p^{2},r)$ has $\#_{\rm t} = q^{2}$ elements in total, out of which there are $\#_{\rm z} = q$ zero-divisors and
$\#_{\rm u} = q^{2} - q = q(q - 1)$ units. Next, let ``$^{\overline{~~}}$" denote reduction modulo $p$; then obviously $\overline{{\cal T}}_{r} = GF(q)$,
the Galois {\it field} of order $q$, and $\overline{\zeta}$ is a primitive element of $GF(q)$. Finally, one notes that any two Galois rings of the
same characteristic and rank are isomorphic.

Now we have a sufficient algebraic background to introduce the
concept of a projective Hjelmslev plane over $GR(p^{2},r)$,
henceforth referred to as $PH(2, q)$.\footnote{This is, of course,
a very specific, and rather elementary, kind of projective
Hjelmslev plane; its most general, axiomatic definition can be
found, for example, in \cite{kli54}--\cite{dem68}.}~ $PH(2, q)$ is
an incidence structure whose points are classes of ordered triples
$(\varrho \breve{x}_1, \varrho \breve{x}_2, \varrho \breve{x}_3)$,
where both $\varrho$ and at least one $\breve{x}_i$ ($i$=1,2,3)
are units, whose lines are classes of ordered triples $(\sigma
\breve{l}_1, \sigma \breve{l}_2, \sigma \breve{l}_3)$, where both
$\sigma$ and at least one $\breve{l}_i$ ($i$=1,2,3) are units, and
the incidence relation is given by
\begin{equation}
\sum_{i=1}^{3} \breve{l}_{i} \breve{x}_{i} \equiv \breve{l}_{1} \breve{x}_{1}  + \breve{l}_{2} \breve{x}_{2} + \breve{l}_{3} \breve{x}_{3} = 0.
\end{equation}
From this definintion it follows that in $PH(2, q)$ --- as in any {\it ordinary} projective plane --- there is a perfect duality between points and lines;
that is, instead viewing the points of the plane as the fundamental entities, and the lines as ranges (loci) of points, we may equally well
take the lines as primary geometric constituents and define points in terms of lines, characterizing a point by the complete set of lines passing
through it.
It is also straightforward to see that this plane contains
\begin{equation}
\#_{\rm trip} = \frac{\#_{\rm t}^{3} -  \#_{\rm z}^{3}}{\#_{\rm u}} = \frac{(q^{2})^{3} -  q^{3}}{q(q - 1)} = \frac{q^{3}(q^{3} - 1)}{q(q - 1)} =
q^{2}\left(q^{2} + q + 1 \right)
\end{equation}
points/lines and that the number of points/lines incident with a given line/point is, in light of Eq.\,(4), equal to the number of non-equivalent couples
$(\varrho \breve{x}_1, \varrho \breve{x}_2)$/$(\sigma \breve{l}_1, \sigma \breve{l}_2)$, i.e.
\begin{equation}
\#_{\rm coup} = \frac{\#_{\rm t}^{2} -  \#_{\rm z}^{2}}{\#_{\rm u}} = \frac{(q^{2})^{2} -  q^{2}}{q(q - 1)} = \frac{q^{2}(q^{2} - 1)}{q(q - 1)} =
q\left(q + 1 \right).
\end{equation}
These figures should be compared with those characterizing ordinary finite planes of order $q$, which read $\overline{\#}_{\rm trip} =
q^{2} + q + 1$ and $\overline{\#}_{\rm coup} =
q + 1$, respectively (e.g., \cite{hir}).

Any projective Hjelmslev plane,  $PH(2, q)$ in particular, is endowed with a very important, and of crucial relevance when it comes to MUBs,
property that has no analogue in an ordinary
projective plane --- the so-called {\it neighbour} (or, as occasionally referred to, non-remoteness) relation. Namely (see, e.g., \cite{kli54}--\cite{dem68}),
we say that two points $A$ and $B$ are neighbour, and write $A \odot B$, if either $A=B$, or $A \neq B$ and there exist two different
lines incident with both; otherwise, they are called nonneighbour, or remote. The same symbol and the dual definition is used for neighbour lines.
Let us find the cardinality of the set of neighbours
of a given point/line of $PH(2, q)$. Algebraically speaking, given a point $\varrho \breve{x}_{i}$, $i$=1,2,3, the points that are its neighbours must
be of the form  $\varrho \left(\breve{x}_{i} + p \breve{y}_{i}\right)$, with $\breve{y}_{i} \in {\cal T}_{r}$; for two points are neighbour iff their
corresponding coordinates differ by a zero divisor \cite{kli54}--\cite{dem68}. Although there are $q^{3}$
different choices for the triple $(\breve{y}_{1}, \breve{y}_{2}, \breve{y}_{3})$, only $q^{3}/q = q^{2}$ of the classes $\varrho \left(\breve{x}_{i}
+ p \breve{y}_{i}\right)$ represent distinct points because $\varrho \left(\breve{x}_{i} + p \breve{y}_{i}\right)$ and
$\varrho \left(\breve{x}_{i} + p (\breve{y}_{i} + \kappa \breve{x}_{i}) \right)$
represent one and the same point as $\kappa$ runs through
all the $q$ elements of ${\cal T}_{r}$. Hence, every point/line of $PH(2, q)$ has $q^{2}$ neighbours, the point/line in question inclusive.
Following the same line of reasoning, but restricting only to couples of coordinates, we find that given a point $P$ and a line ${\cal L}$,
$P$ incident with ${\cal L}$, there exist exactly $\left( q^{2}/q = \right)q$ points on ${\cal L}$ that are neighbour to $P$ and, dually,
$q$ lines through $P$ that are neighbour to ${\cal L}$.

Clearly, as $A \odot A$ (reflexivity), $A \odot B$ implies $B \odot A$ (symmetry) and $A \odot B$ and $B \odot C$ implies
$A \odot C$ (transitivity), the neighbour relation is an {\it equivalence} relation. Given ``$\odot$" and a point $P$/line ${\cal L}$, we call
the subset of all points $Q$/lines ${\cal K}$ of $PH(2, q)$ satisfying $P \odot Q$/${\cal L} \odot {\cal K}$ the neighbour class of $P$/${\cal L}$.
And since ``$\odot$" is an equivalence relation, the aggregate of neighbour classes {\it partitions} the plane, i.e. the plane consists of a {\it dis}joint
union of neighbour classes of points/lines. The modulo-$p$-mapping  then ``induces" a so-called canonical epimorphism of $PH(2, q)$
into $PG(2, q)$, the ordinary projective plane defined over  $GF(q)$, with the neighbour classes being the cosets of this epimorphism \cite{dem68}.
Loosely rephrased, $PH(2, q)$ comprises $q^{2} + q + 1$ ``clusters" of neighbour points/lines, each of cardinality $q^{2}$, such that its restriction
modulo the neighbour relation is the ordinary projective plane $PG(2, q)$ every single point/line of which encompasses the whole ``cluster" of
these neighbour points/lines. Analogously, each line of $PH(2, q)$ consists of $q+1$ neighbour classes, each of cardinality $q$, such that its
``$^{\overline{~~}}$" image is the ordinary projective line in the $PG(2, q)$ whose points are exactly these neighbour classes.
\begin{figure}[t]
\centerline{\includegraphics[width=8truecm,clip=]{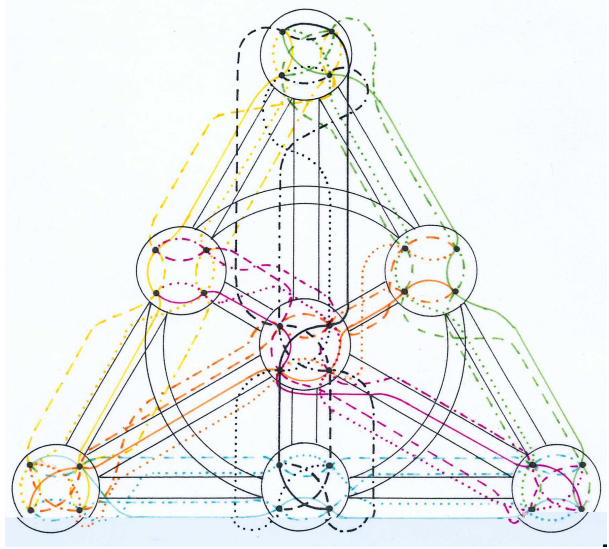}}
\caption{A schematic sketch of the structure of the simplest
projective Hjelmslev plane, $PH(2,2)$. Shown are all the 28 of its
points (represented by small filled circles), grouped into seven
pairwise disjoint sets (neighbour classes), each of cardinality
four, as well as 24 of its lines (drawn as solid, dashed, dotted
and dot-dashed curves), forming six different neighbour classes.
In order to make the sketch more illustrative, different neighbour
classes of lines have different colour. Also shown is the
associated Fano plane, $PG(2,2)$, whose points are represented by
seven big circles, six of its lines are drawn as pairs of line
segments and the remaining line as a pair of concentric circles.
Notice the intricate character of pairwise intersection of the
lines of $PH(2,2)$; two lines from distinct neighbour classes have
just one point in common, while any two lines within a neighbour
class share ($q$=)2 points, both of the same neighbour class.}
\end{figure}

Let us illustrate these remarkable properties on the simplest possible example that is furnished by $PH(2, q=2)$, i.e. the plane defined
over $GR(4,1)$ whose epimorphic ``shadow" is the simplest projective plane $PG(2, 2)$ --- the Fano plane. As partially depicted
in Fig.\,1, this plane consists of seven classes of quadruples of neighbour points/lines, each point/line featuring three classes of couples
of neighbour lines/points. When modulo-two-projected, each quadruple of neighbour points/lines goes into a single point/line of the
associated Fano plane.

The most relevant geometrical object for our model \cite{sp05} is, of course, a {\it conic}, that is a curve ${\cal Q}$ of $PH(2, q)$ whose points obey
the equation
\begin{equation}
{\cal Q}:~ \sum_{i \leq j} c_{ij} \breve{x}_{i} \breve{x}_{j} \equiv c_{11} \breve{x}_{1}^{2} + c_{22} \breve{x}_{2}^{2} +
c_{33} \breve{x}_{3}^{2} + c_{12} \breve{x}_{1} \breve{x}_{2} + c_{13} \breve{x}_{1} \breve{x}_{3} + c_{23} \breve{x}_{2} \breve{x}_{3} = 0,
\end{equation}
with at least one of the $c_{ij}~'s$ being a unit of $GR(p^{2},r)$. In particular, we are interested in a {\it proper} conic, which is a conic
whose equation cannot be reduced into a form featuring fewer variables whatever coordinate transformation one would employ. It is
known (see, e.g., \cite{chj}) that the equation of a proper conic of $PH(2, q)$ can always be brought into a ``canonical" form
\begin{equation}
{\cal Q}^{\star}:~ \breve{x}_{1} \breve{x}_{3} - \breve{x}_{2}^{2} = 0
\end{equation}
from which it readily follows that any such conic is endowed, like a line, with $q^{2} + q = q(q + 1)$ points; $q^{2}$ of them are of the form
\begin{equation}
\varrho \breve{x}_{i} = (1, \sigma, \sigma^{2}),
\end{equation}
where the parameter $\sigma$ runs through all the elements of $GR(p^{2},r)$, whilst
the remaining $q$ are represented by
\begin{equation}
\varrho \breve{x}_{i} = (0, \delta, 1),
\end{equation}
with $\delta$ running through all the zero-divisors of $GR(p^{2},r)$.
And each point of a proper conic, like that of a line, has $q$ neighbours; for the neighbours of a particular point $\sigma = \sigma_{0}$
of (9) are of the form
\begin{equation}
\varrho \breve{x}_{i} = (1, \sigma_{0} + p \kappa, (\sigma_{0} + p \kappa)^{2}) = (1, \sigma_{0} + p \kappa, \sigma_{0}^{2} + p 2 \kappa)
\end{equation}
and there are obviously $q$ of them (the point in question inclusive) as $\kappa$ runs through  ${\cal T}_{r}$, and all
the $q$ points of (10) are the neighbours of any of them. All in all, a proper conic, like a line, of $PH(2, q)$ features $q + 1$ pairwise
disjoint classes of neighbour points, each having $q$ elements, these classes being the single points of its modular image
in $PG(2, q)$. To illustrate the case, several proper conics in $PH(2,2)$ are shown in Fig.\,2.
\begin{figure}[t]
\centerline{\includegraphics[width=9truecm,clip=]{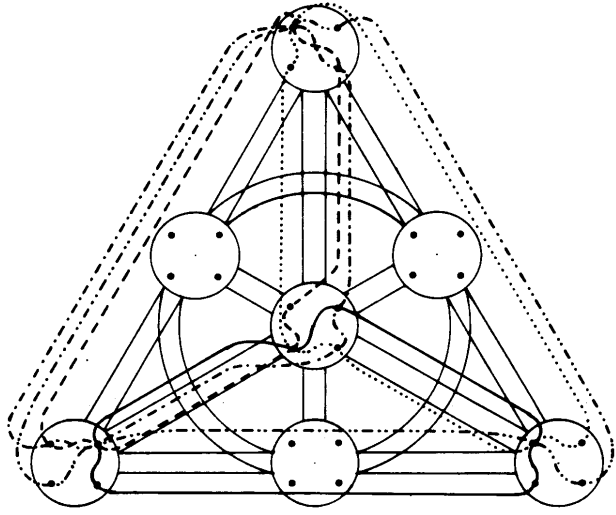}}
\caption{The forms of five different proper conics located in $PH(2,2)$: $\breve{x}_{1} \breve{x}_{3} - \breve{x}_{2}^{2} = 0$ (solid curve),
$\breve{x}_{1} \breve{x}_{2} - \breve{x}_{3}^{2} = 0$ (dashed), $\breve{x}_{2} \breve{x}_{3} - \breve{x}_{1}^{2} = 0$ (dotted),
$\breve{x}_{1} \breve{x}_{2} + \breve{x}_{3}^{2} = 0$ (dot-dashed) and $\breve{x}_{1} \breve{x}_{2} + \breve{x}_{1} \breve{x}_{3} +
\breve{x}_{2} \breve{x}_{3} = 0$ (dash-doubledotted). Note the intricacies of pairwise intersections between the conics.}
\end{figure}

At this point our algebraic geometrical machinery is elaborate enough to generalize and qualitatively complete
the geometrical picture of MUBs proposed in \cite{sp05} where we have argued that a basis of ${\cal H}_q$, $q$ given by (3), can be
regarded as a point of an arc in $PG(2, q)$, with a
complete set of MUBs corresponding to a proper conic (or, in the case of $p$=2, to a more general geometrical object called oval).
This model, however, lacks a geometrical interpretation of the individual vectors of a basis, which can be achieved in our extended
projective setting {\it \` a la} Hjelmslev only. Namely, taking any complete, i.e. of cardinality $q$+1, set of MUBs, its {\it bases} are now viewed as
the {\it neighbour classes} of points of a proper conic of $PH(2, q)$ and the {\it vectors} of a given basis  have their counterpart in the {\it points}
of the corresponding neighbour class. The property of different  vectors of a basis being pairwise {\it orthogonal} is then geometrically embodied
in the fact that the corresponding points are all {\it neighbour}, whilst the property of two different bases being {\it mutually unbiased}
answers to the fact that the points of any two neighbour classes are {\it remote} from each other. It is left to the reader as an easy exercise
to check that ``rephrasing these statements modulo $p$" one recovers all the conic-related properties of MUBs given in  \cite{sp05},
irrespective of the value of $p$. The ($p$=2) case of ``non-conic" MUBs is here, however, much more complex and intricate than that in the
ordinary projective planes and will properly be dealt with in a separate paper.

To conclude, it must be stressed that this remarkable analogy between complete sets of MUBs and ovals/conics is worked out at the level of cardinalities only
and thus still remains a conjecture. Hence, the next crucial step
to be done is to construct an expliciting mapping by associating a MUB to each neighbour class of the points of the conic and a state vector of this MUB to a particular point
of the class. This is a much more delicate issue, as there are (at least) two non-isomorphic kinds of projective Hjelmslev planes of order $q = p^r$ that have exactly the same
``cardinality" properties, viz. the plane defined over the Galois ring $GR(p^2,r)$ and the one defined over the ring of ``dual" numbers,  $GF(q)[x]/(x^2) \cong GF(q) + eGF(q)$, where
$e^2 = 0$. Even for the simplest case ($p$=2 and $r$=1) there is an intricate difference in geometry between the two planes, as the former contains ($q^2+q+1=$)7-arcs,
while the latter not (see, e.g., \cite{hl}). A thorough exploration of the fine structure of these two Hjelmslev geometries, as well as of a number of other finite Hjelmslev and
related ring planes,
is therefore a principal theoretical task for making further progress in this direction.

\vspace*{.5cm}
\noindent
{\bf Acknowledgement}\\
The first author is grateful to Prof. Olav Arnfinn Laudal for a
number of enlightening comments concerning the structure of
projective geometries over (Galois) rings and to Prof. Mark
Stuckey for being a great help to him in obtaining a copy of
\cite{kle59}. We also thank two of the referees for their
constructive comments and Mr. Pavol Bend\'{\i}k for careful
drawing of the figures.

\end{document}